\documentclass[11pt,a4paper]{article}
\usepackage{amsmath,amssymb,color,enumerate,multirow,url,xcolor}

\newcommand{\AB}{\mathit{AB}}
\newcommand{\ABB}{\mathit{ABB}}
\newcommand{\ab}{\mathit{ab}}
\newcommand{\abb}{\mathit{abb}}
\newcommand{\mq}[1]{\left< #1 \right>}
\newcommand{\mql}[1]{\left< #1 \right.}
\newcommand{\mqr}[1]{\left. #1 \right>}
\newcommand{\case}[1]{\par\noindent\textbf{#1.}}
\newcommand{\subcase}[1]{\par\noindent\textit{#1.}}

\newtheorem{theorem}{Theorem}

\title{The Worst Case Number of Questions\\
in Generalized AB Game with and without White-peg Answers}

\author{
Gerold J\"ager\footnote{
Department of Mathematics and Mathematical Statistics,
University of Ume{\aa},
SE-901-87 Ume{\aa}, Sweden,
\texttt{e-mail:\ gerold.jaeger@math.umu.se}}
\and
Marcin Peczarski\footnote{
Institute of Informatics, University of Warsaw,
ul.\ Banacha 2, PL-02-097 Warszawa, Poland,
\texttt{email:\ marpe@mimuw.edu.pl}}
}

\date{Research done between November 2011 and July 2012}

\begin{document}
\maketitle

\begin{abstract}
\noindent The AB game is a two-player game, where the codemaker has to
choose a secret code and the codebreaker has to guess it in as few
questions as possible. It is a variant of the famous Mastermind game,
with the only difference that all pegs in both, the secret and the
questions must have distinct colors. In this work, we consider the
Generalized AB game, where for given arbitrary numbers $p$, $c$ with
$p \le c$ the secret code consists of $p$ pegs each having one of $c$
colors and the answer consists only of a number of black and white
pegs. There the number of black pegs equals the number of pegs
matching in the corresponding question and the secret in position and
color, and the number of white pegs equals the additional number of
pegs matching in the corresponding question and the secret only in
color. We consider also a variant of the Generalized AB game, where
the information of white pegs is omitted. This variant is called
Generalized Black-peg AB game. Let $\ab(p,c)$ and $\abb(p,c)$ be the
worst case number of questions for Generalized AB game and Generalized
Black-peg AB game, respectively. Combining a computer program with
theoretical considerations, we confirm known exact values of
$\ab(2,c)$ and $\ab(3,c)$ and prove tight bounds for $\ab(4,c)$.
Furthermore, we present exact values for $\abb(2,c)$ and $\abb(3,c)$
and tight bounds for $\abb(4,c)$.
\end{abstract}

\section{Introduction}

The AB game is a variant of the famous Mastermind game, which has
attracted much attention in literature in the longer and recent past.
Mastermind leads to a rich source of recreational \cite{Knu76,KL93}
and combinatorial open problems \cite{Ch83}. Recently, theoretical
results considering the hardness of Mastermind have been presented
\cite{DW12,SZ06,Vig12}. On the other hand, there are also many
interesting applications of Mastermind, e.g., in cryptography
\cite{FL12} and bioinformatics \cite{GET11}. Most research has been
done on the expected-case and worst-case behavior of Mastermind
strategies, e.g., \cite{CLH07}. In this context also methods from
completely different fields have shown to be helpful, e.g., graph
partitioning \cite{CLH04} and evolutionary algorithms \cite{GCG09}.

Many variants of Mastermind have been considered, e.g., Black-peg
Mastermind \cite{Good09,JP11} and static Mastermind \cite{God03}.
Another variant of Mastermind is the AB game, which is the topic of
this work and which has already been considered in \cite{CL04,HL09}.
It is also known as ``bulls and cows'' game. Whereas the most popular
version of the AB game is played with $4$ pegs and $10$ colors, the
Generalized AB game is played with $p$ pegs and $c$ colors, where $c
\ge p$. We denote this game by $\AB(p,c)$. Two players are involved in
this game, which are called the \emph{codemaker} and the
\emph{codebreaker}. In the beginning of the game, the codemaker
chooses a secret containing $p$ pegs, each of different color. The
codebreaker tries to identify the secret by asking questions which
also contain $p$ pegs, each of different color. The codemaker answers
the questions using black and white pegs. The number of black pegs
informs, how many pegs in the question match pegs in the secret in
position and color. The number of white pegs gives the information,
how many further pegs in the question match pegs in the secret only in
color, but not in position. The goal of the codebreaker is to minimize
the number of questions needed to guess the secret. The game ends when
a question is answered with $p$ black pegs, where this last question
is counted to the total number of asked questions. Note that the only
difference to Mastermind is that in the AB game all pegs in both, the
secret and the questions must have distinct colors. Generalized
Black-peg AB game, denoted as $\ABB(p,c)$, is a modification of the AB
game, where the answers contain only black pegs. This modification is
analogous to the modification of Mastermind to Black-peg Mastermind
\cite{JP11}. We denote by $\ab(p,c)$ and $\abb(p,c)$ the worst case
number of questions in the $\AB(p,c)$ and $\ABB(p,c)$ game,
respectively. If the game has $c$ colors, we name the colors using
consecutive numbers: $0, 1, \dots, c-1$. Analogously, if the game has
$p$ pegs, we name the pegs using consecutive numbers: $0, 1, \dots,
p-1$. We denote a question by $\mq{k_0, k_1, \dots, k_{p-1}}$, where
the color $k_i$ is asked at position $i$ for $i=0,1,\dots,p-1$.

It has been proved in \cite{CL04} that
\begin{equation}
\label{eq:ab2}
\ab(2,c) = \lceil c/2 \rceil + 1
  = \lfloor (c+1)/2 \rfloor + 1, \quad\textnormal{if $c \ge 2$}.
\end{equation}
and in \cite{HL09} that
\begin{equation}
\label{eq:ab3}
\ab(3,c) = \left\{
  \begin{array}{l l}
    \lfloor c/3 \rfloor + 3,     & \textnormal{if $3 \le c \le 7$,}\\
    \lfloor (c+1)/3 \rfloor + 3, & \textnormal{if $c \ge 8$.}\\
  \end{array} \right.
\end{equation}
We agree that the above formula is correct. However, we think that the
proof given in \cite{HL09} is wrong or at least not complete. In
particular, it is not well justified that the state after the
\emph{structural reduction} is \emph{as hard as or harder} than the
initial state \cite[p. 173, the last par.]{HL09}.

We consider the $\AB$ game in Section \ref{sec:ab}. We prove equations
(\ref{eq:ab2}) and (\ref{eq:ab3}) independently using the approach
introduced in \cite{JP09} and then extended in \cite{JP11} (see
Sections \ref{sec:ab2} and \ref{sec:ab3}). Compared to \cite{CL04} and
\cite{HL09}, where different methods are proposed, our proof benefits
from the same auxiliary results (see Section \ref{sec:abaux}).
Furthermore, it is much simpler and only needs to check some values
received by a computer program. Moreover, our approach allows us to
give in Section \ref{sec:ab4} a similar result for four pegs, namely
Theorem~\ref{thm:ab4}.
\begin{theorem}\label{thm:ab4}
It holds:
{\setlength\arraycolsep{1.7pt}
\begin{eqnarray}
\label{eq:ab4eq}\ab(4,c) & = & \left\{
  \begin{array}{ll}
    \lfloor (c+2)/3 \rfloor + 3, & \textnormal{if $4 \le c \le 11$},\\
    8                          , & \textnormal{if $c = 12,13$},\\
  \end{array} \right.
\\[\parskip]
\label{eq:ab4lower}\ab(4,c) & \ge & \lfloor (c+3)/4 \rfloor + 4,
  \quad\textnormal{if $c \ge 14$},
\\[\parskip]
\label{eq:ab4upper}\ab(4,c) & \le & \lfloor (c+3)/4 \rfloor + 5,
  \quad\textnormal{if $c \ge 14$}.
\end{eqnarray}}
\end{theorem}

We close Section \ref{sec:ab} with some considerations about the value
of $\ab(p,p)$ in Section \ref{sec:abeq}.

The whole Section \ref{sec:abb} is devoted to the lower and upper
bounds for the worst case number of questions in the $\ABB$ game.
We receive Theorem \ref{thm:abb}.
\begin{theorem}\label{thm:abb}
It holds:
{\setlength\arraycolsep{1.7pt}
\begin{eqnarray}
\label{eq:abb2}\abb(2,c) & = & c, \phantom{\null+1}\quad
               \textnormal{if $c \ge 2$},\\[\parskip]
\label{eq:abb3}\abb(3,c) & = & c+1, \quad
               \textnormal{if $c \ge 3$},\\[\parskip]
\label{eq:abb4lower}\abb(4,c) & \ge & c+1, \quad
                    \textnormal{if $c \ge 4$},\\[\parskip]
\label{eq:abb4upper}\abb(4,c) & \le & \left\{
  \begin{array}{ll}
    c+1, & \textnormal{if $4 \le c \le 10$},\\
    c+2, & \textnormal{if $c \ge 11$}.\\
  \end{array} \right.
\end{eqnarray}}
\end{theorem}

The results presented in this paper are obtained with the help of a
computer program, which is a modification of the program used in our
previous papers about Mastermind \cite{JP09, JP11}. A compressed
archive with the complete source code of the program and scripts for
reproducing all computations can be downloaded from \cite{ABcode}.

\section{AB Game with White Pegs in Answers}\label{sec:ab}

We verified equations (\ref{eq:ab2}) and (\ref{eq:ab3}) for $c\le13$,
using the computer program. Additionally, we computed the values for
$p=4$. The results are presented in Table \ref{tbl:ab}. We have
adapted the approach introduced in \cite{JP09} and extended in
\cite{JP11} to obtain formulas for an arbitrary number of colors. As
previously, we introduce two auxiliary games: $\AB_*$ and $\AB^*$.

\begin{table}[htb]
\caption{Computed values $\ab(p,c)$ for $2 \le p \le 4$ and
         $p \le c \le 13$}
\label{tbl:ab}
\begin{center}
\begin{tabular}{|cc|*{12}{r}|}
\hline
&&\multicolumn{12}{c|}{$c$}\\
&   & 2 & 3 & 4 & 5 & 6 & 7 & 8 & 9 &10 &11 &12 &13 \\
\hline
\multirow{3}{0.5em}{$p$}
& 2 & 2 & 3 & 3 & 4 & 4 & 5 & 5 & 6 & 6 & 7 & 7 & 8 \\
& 3 &   & 4 & 4 & 4 & 5 & 5 & 6 & 6 & 6 & 7 & 7 & 7 \\
& 4 &   &   & 5 & 5 & 5 & 6 & 6 & 6 & 7 & 7 & 8 & 8 \\
\hline
\end{tabular}
\end{center}
\end{table}

The $\AB_*(p,c)$ game is the $\AB(p,c)$ game in which an additional
color $c$ in questions is allowed. Note that the additional color
cannot appear in a secret and that the additional color can appear in
a question more than twice, but all other ``normal'' colors only once.
As asking a question containing $p$ times the additional color does
not makes any sense, we do not need to consider such questions in the
computer program.

The $\AB^*(p,c,x)$ game, where $x \in \mathbb{N}_0$ with $px \le c$,
is the $\AB(p,c)$ game, where the beginning $x$ questions are fixed,
namely question $k$ is $\mql{pk}, pk+1, $ $ \dots,
\mqr{pk+p-1}$ for $k=0,1,\dots,x-1$. Note that the $\AB^*(p,c,0)$
game is equivalent to the $\AB(p,c)$ game. We denote by $\ab_*(p,c)$
and $\ab^*(p,c,x)$ the worst case number of questions in the
$\AB_*(p,c)$ and $\AB^*(p,c,x)$ game, respectively.

\subsection{Auxiliary Results}\label{sec:abaux}

Every strategy for $\AB(p,c)$ using at most $q$ questions allows to
win $\AB_*(p,c)$ using also at most $q$ questions. Hence, we have
$\ab(p,c) \ge \ab_*(p,c)$. We can easily transform a strategy for
$\AB_*(p,c+1)$ into a strategy for $\AB_*(p,c)$ by changing the
additional color $c+1$ of the $\AB_*(p,c+1)$ game into the color $c$
which is unused in secrets of the $\AB_*(p,c)$ game and plays the role
of an additional color in $\AB_*(p,c)$. Hence, $\ab_*(p,c+1) \ge
\ab_*(p,c)$. If we ask the first question containing $m \le p$
``normal'' colors in the $\AB_*(p,c)$ game and the adversary gives us
the empty answer, we are forced to play the $\AB_*(p,c-m)$ game.
Hence, we have
\begin{displaymath}
\ab_*(p,c) \ge 1+\min_{1 \le m \le p} \ab_*(p,c-m) = 1+\ab_*(p,c-p).
\end{displaymath}
Consequently for $c \ge c_0$ we have
\begin{equation}
\label{eq:ablowerbound}
\ab(p,c) \ge \lfloor (c-c_0)/p \rfloor + \ab_*(p,c_0).
\end{equation}

On the other hand, every strategy for $\AB^*(p,c,x)$ is a proper
strategy for $\AB(p,c)$. Hence, we have $\ab(p,c) \le \ab^*(p,c,x)$.
Now, let the number of colors be $c=px+m$, where $x \ge p$ and
$m\in\mathbb{N}_0$. We consider a strategy for the $\AB^*(p,px+m,x)$
game. There are at least $x-p$ empty answers among the first $x$
questions which discard $(x-p)p$ colors. Hence, the game is reduced to
the $\AB^*(p,p^2+m,p)$ game and we have
\begin{equation}
\label{eq:abupperbound}
\ab(p,px+m) \le \ab^*(p,px+m,x) \le x-p + \ab^*(p,p^2+m,p).
\end{equation}

In the following subsections we combine the inequalities
(\ref{eq:ablowerbound}), (\ref{eq:abupperbound}) with computed values
to obtain lower and upper bounds for the worst case number of
questions in the $\AB(p,c)$ game for a fixed number of pegs and an
arbitrary number of colors.

\subsection{Two Pegs}\label{sec:ab2}

For two pegs the computer program yields $\ab_*(2,3)=3$ and
$\ab^*(2,5,2)=\ab^*(2,6,2)=4$.
Using equation (\ref{eq:ablowerbound}) for $c_0=3$, we receive
for $c \ge 3$ that
\begin{displaymath}
\ab(2,c) \ge \lfloor(c-3)/2\rfloor + 3 =\lfloor(c+1)/2\rfloor + 1.
\end{displaymath}
Moreover, by equation (\ref{eq:abupperbound}), for $x \ge 2$ and
$m=1,2$ we have
\begin{displaymath}
\ab(2,2x+m) \le x-2 + \ab^*(2,4+m,2) = x+2,
\end{displaymath}
which implies for $c \ge 5$ that it holds
\begin{displaymath}
\ab(2,c) \le \lfloor (c+1)/2 \rfloor + 1.
\end{displaymath}
The above inequalities together with the values from Table
\ref{tbl:ab} confirm equation (\ref{eq:ab2}).

\subsection{Three Pegs}\label{sec:ab3}

For three pegs the computer program yields $\ab_*(3,8)=6$ and
$\ab^*(3,14,3)=\ab^*(3,15,3)=\ab^*(3,16,3)=8$. Using equation
(\ref{eq:ablowerbound}) for $ c_0=8$, we receive for $c \ge 8$ that
\begin{displaymath}
\ab(3,c) \ge \lfloor(c-8)/3\rfloor + 6 = \lfloor(c+1)/3\rfloor + 3.
\end{displaymath}
Moreover, by equation (\ref{eq:abupperbound}), for $x \ge 3$ and
$m=5,6,7$ we have
\begin{displaymath}
\ab(3,3x+m) \le x-3 + \ab^*(3,9+m,3) = x+5,
\end{displaymath}
which implies for $c \ge 14$ that it holds
\begin{displaymath}
\ab(3,c) \le \lfloor (c+1)/3 \rfloor + 3.
\end{displaymath}
The above inequalities together with the values from Table
\ref{tbl:ab} confirm equation (\ref{eq:ab3}).

\subsection{Four Pegs}\label{sec:ab4}

For $p=4$ we cannot give an exact formula, but we present close lower
and upper bounds, where the gap between the bounds does not exceed
one question.

The program yields $\ab_*(4,13) \ge 8$. Using equation
(\ref{eq:ablowerbound}) for $c_0=13$, we receive for $c \ge 13$ that
\begin{displaymath}
\ab(4,c) \ge \lfloor(c-13)/4\rfloor + 8 = \lfloor(c+3)/4\rfloor + 4,
\end{displaymath}
which confirms inequality (\ref{eq:ab4lower}).
By the computer program,
$\ab^*(4,17,4) \le 10$,
$\ab^*(4,18,4) \le 10$,
$\ab^*(4,19,4) \le 10$,
$\ab^*(4,20,4) \le 10$ hold.
Note that in these cases we do not know the exact values, but
only upper bounds.
Using equation (\ref{eq:abupperbound}) for $x \ge 4$ and
$m=1,2,3,4$ we have
\begin{displaymath}
\ab(4,4x+m) \le x-4 + \ab^*(4,16+m,4) \le x+6,
\end{displaymath}
which implies that it holds for $c \ge 17$
\begin{displaymath}
\ab(4,c) \le \lfloor (c-1)/4 \rfloor + 6 = \lfloor (c+3)/4 \rfloor + 5.
\end{displaymath}
We computed directly upper bounds for the three missing
values, namely 14, 15 and 16 colors. The program returned the
bounds $\ab(4,14) \le 9$, $\ab(4,15) \le 9$ and $\ab(4,16) \le 9$.
This closes the proof of inequality (\ref{eq:ab4upper}).
Table~\ref{tbl:ab} contains the values up to 13 colors, which confirms
equation (\ref{eq:ab4eq}).

\subsection{Equal Number of Pegs and Colors}\label{sec:abeq}

The games $\AB(p,p)$ and $\ABB(p,p)$ are equivalent, as the equality
$p=c$ implies $b+w=p$, where $b$ is the number of black pegs in the
answer and $w$ is the number of white pegs in the answer. Hence, if
the number of colors is equal to the number of pegs, $w$ is uniquely
determined by $b$. Therefore, it holds $\ab(p,p) = \abb(p,p)$, which
is the motivation to consider the only-black-peg version of the game.

The lower bound $\ab(p,p) = \Omega(p)$ has been proved in \cite{KT86}.
This result can be reformulated as follows.
We have $p!$ possible secrets. There are $c$
possible answers to each question, namely the number of black pegs
could be $0$, $1$, $2$, $\dots$, $p-2$, $p$. Note that the answer
$p-1$ black pegs is not possible. As the answer $p$ black pegs
finishes the game, for every question we have at most $p-1$ possible
continuations of the game. Therefore, if $p>2$ and we ask $q$
questions, we can solve at most
\begin{displaymath}
T(p,q) = \sum_{i=0}^{q-1}(p-1)^i = \frac{(p-1)^q-1}{p-2} < p^q
\end{displaymath}
secrets and it must hold $p! \le T(p,q)$.
Note that $T(2,q)=q$.
Using Stirling's approximation $p! > (p/e)^p$,
we obtain an asymptotic lower bound
\begin{displaymath}
\ab(p,p) > p\left(1-\frac{1}{\ln p}\right).
\end{displaymath}

\begin{table}[htb]
\caption{Computed values for $2 \le p \le 6$}
\label{tbl:abpp}
\begin{center}
\begin{tabular}{|c|*{5}{r}|}
\hline
$p$           & 2 & 3 & 4 & 5 & 6 \\
\hline
$\ab(p,p)$    & 2 & 4 & 5 & 6 & 7 \\
\hline
$q_{\min}(p)$ & 2 & 3 & 4 & 5 & 5 \\
\hline
\end{tabular}
\end{center}
\end{table}

The upper bound $\ab(p,p) = O(p \log p)$ has been shown in \cite{KT86}.
% and has been improved by a constant factor in \cite{OS13}.
% \COMM{We can remove the last citation if you wish or cite more results
% from this work, i.e., $abb(p,c) \le p \log p + p + c$ for $c > p$.}

Table~\ref{tbl:abpp} contains exact values for $\ab(p,p)$, computed by
the program, in the second row. The last row contains the smallest
value of $q$ satisfying the inequality $p! \le T(p,q)$, which gives a
lower bound for $\ab(p,p)$.

\section{AB Game without White Pegs in Answers}\label{sec:abb}

\subsection{Lower Bounds}

We prove lower bounds of the $\ABB(p,c)$ game by showing a
counterstrategy for the codemaker. The counterstrategy is parametrized
with two numbers $q,r\in\mathbb{N}$, where $r \ge p$ and these
parameters depend on $p$, but not on $c$.

The counterstrategy starts with the initial phase, where the codemaker
answers the first $c-r$ questions with zero black pegs. This strategy
is valid, as after that at each peg position there are at least $r$
possible colors. If the codemaker chooses an arbitrary color for the
first peg, and an arbitrary unused color for the following pegs, then
this process leads to a possible secret which would receive the answer
of zero black pegs in the $c-r$ questions. On the other hand, it is
not always possible for the codemaker to answer the first $c-p+1$ (or
more) questions with zero black pegs. This can be seen by the
following example.

\case{Example} Consider the game $\ABB(4,7)$, and let the codebreaker
ask the $c-p+1=4$ questions: $\mq{0,1,2,3}$, $\mq{1,2,3,0}$,
$\mq{2,3,0,1} $, $\mq{3,0,1,2}$. If the codemaker would answer all
these questions with zero black pegs, then the only possible colors
for a secret would be $4$, $5$, $6$, but no secret exists with $4$
pegs and only $3$ different colors.

After the initial phase an end-game is played, where the goal of the
codemaker is to force the codebreaker to ask more than $q$ questions.
To ease the analysis of the end-game, we transform the set of possible
secrets, but we define only transformation rules which do not increase
the worst case number of questions in the end-game. As some
transformation rules change colors, they also affect the set of
questions. To overcome this problem, we extend the set of allowed
questions. The codebreaker is not restricted to ask only questions
with distinct colors in the end-game. Although extending the set of
questions could decrease the worst case number of questions required
to win the end-game, by choosing a suitable value of the
counterstrategy parameter $r$ we receive the desired tight lower
bounds.

After answering the $c-r$ questions with zero black pegs, some colors
are excluded from being present at some positions in the secret. For
every peg position, we consider a set of possible colors for that
position. The cardinality of that set is at least $r$. The sequence of
such sets for all positions is called an \emph{end-game state} or
simply a \emph{state} for short. We represent the state by a table
containing $p$ rows. The row $i$ contains the colors which are still
possible at peg position $i$. In the following, we denote for a given
color the set of row numbers of the state, where this color appears
in, as its \emph{row set}. We denote a row set of cardinality $1$,
$2$, $3$ or $4$ as \emph{single row set}, \emph{pair row set},
\emph{triple row set} and \emph{quadruple row set}, respectively.

Below we formally write all state transformation rules.
An application example is shown in Figure \ref{fig:abbl:rulesex}.

\subcase{Rule 1}
Any color can be removed from any row.

\subcase{Rule 2}
Colors can be permuted.

\subcase{Rule 3}
Rows can be permuted.

\subcase{Rule 4}
If the colors $k_1$ and $k_2$ have disjoint row sets, then the color
$k_2$ can be replaced by the color $k_1$.

\begin{figure}[t]
\begin{center}
\begin{displaymath}
\begin{array}{c}
\left(
\begin{array}{ccccc}
  0 & 1 & 2 & 3 &   \\
  4 & 5 & 6 & 7 & 8 \\
  9 & 10 & 11 & 12 & \\
\end{array}
\right)
\mathrel{\mathop{\kern0pt{\hbox to40pt{\rightarrowfill}}}
\limits^{\textnormal{Rule 1}}}
\left(
\begin{array}{cccc}
  0 & 1 & 2 & 3 \\
  4 & 5 & 6 & 7 \\
  9 & 10 & 11 & 12 \\
\end{array}
\right)
\mathrel{\mathop{\kern0pt{\hbox to40pt{\rightarrowfill}}}
\limits^{\textnormal{Rule 2}}}
\\[1cm]
\left(
\begin{array}{cccc}
  0 & 1 & 2 & 3 \\
  4 & 5 & 6 & 7 \\
  8 & 9 & 10 & 11 \\
\end{array}
\right)
\mathrel{\mathop{\kern0pt{\hbox to40pt{\rightarrowfill}}}
\limits^{\textnormal{Rule 4}}
\limits_{\substack{
  k_1=0 \\
  k_2=4
}}}
\left(
\begin{array}{cccc}
  0 & 1 & 2 & 3 \\
  0 & 5 & 6 & 7 \\
  8 & 9 & 10 & 11 \\
\end{array}
\right)
\mathrel{\mathop{\kern0pt{\hbox to40pt{\rightarrowfill}}}
\limits^{\textnormal{Rule 4}}
\limits_{\substack{
  k_1=0 \\
  k_2=8
}}}
\\[1cm]
\left(
\begin{array}{cccc}
  0 & 1 & 2 & 3 \\
  0 & 5 & 6 & 7 \\
  0 & 9 & 10 & 11 \\
\end{array}
\right)
\mathrel{\mathop{\kern0pt{\hbox to40pt{\rightarrowfill}}}
\limits^{\textnormal{Rule 4}}}
\cdots
\mathrel{\mathop{\kern0pt{\hbox to40pt{\rightarrowfill}}}
\limits^{\textnormal{Rule 4}}}
\left(
\begin{array}{cccc}
  0 & 1 & 2 & 3 \\
  0 & 1 & 2 & 3 \\
  0 & 1 & 2 & 3 \\
\end{array}
\right)
\end{array}
\end{displaymath}
\caption{Rule application example for $p=3$}
\label{fig:abbl:rulesex}
\end{center}
\end{figure}

Rule 1 is correct, as the set of possible secrets is not increased by
omitting a color for a fixed peg. However, we cannot remove too many
colors, because this would result in decreasing the worst case number
of questions. It is also clear that Rules 2 and 3 are correct, as
they do not change the worst case number of questions.

The proof of Rule 4 is more complicated. Let $S_1$ and $S_2$ be states
before and after applying Rule 4, respectively, and let $R_1$ and
$R_2$ be the row sets of the colors $k_1$ and $k_2$, respectively. We
need to show the implication that if the codebreaker can win $S_1$ in
$q$ questions, then he or she can win $S_2$ also in $q$ questions. Let
the codebreaker have a $q$-question winning strategy $X_1$ for $S_1$.
We construct a strategy $X_2$ allowing the codebreaker to win $S_2$ in
$q$ questions. We replace in $X_1$ the color $k_2$ by the color $k_1$
in all secrets. We exchange in $X_1$ the colors $k_1$ and $k_2$ in all
questions, but only at positions which are in the row set $R_2$. The
assumption that the row sets $R_1$ and $R_2$ are disjoint is
important, because it implies that the secret distinctness is
preserved and then answers are preserved. Formally, if in $X_1$ the
question $q_1$ answers the secret $s_1$ with $b$ black pegs, $q_1$ is
mapped to $q_2$, and $s_1$ is mapped to $s_2$, then in $X_2$ the
question $q_2$ answers the secret $s_2$ also with $b$ black pegs. The
questions in $X_2$ remain valid, because we allowed the codebreaker to
ask all combinations of colors. Some secrets, namely those containing
the colors $k_1$ and $k_2$ in $S_1$, become not valid in $S_2$,
because in $S_2$ they contain two times the color $k_1$. This causes
no problems, as by omitting these secrets the worst case number of
questions can only become smaller, but not larger. Finally, the
transformation described by Rule 4 is an onto function, i.e., if $s_2$
is a valid secret in $S_2$, then there must be a valid secret $s_1$ in
$S_1$, such that Rule 4 maps $s_1$ to $s_2$.

Now, to prove the lower bound for a given $p$, we consider all
possible states and we apply the above rules to them. The goal is to
reduce all states to a small set of non-reducible ones. The number of
these states and the states itself must not depend on $c$. We leave
exactly $r$ colors for each row, using Rule 1. After that we eliminate
all disjoint row sets by Rule 4. As all rows contain the same number
of colors, this will also eliminate all single row sets. Because of
Rule 2, we can assume that the state contains exactly the colors $0$,
$1$, $\dots$, $c_0$. Rule 3 is used to throw out isomorphic states. As
the colors $c_0+1$, $c_0+2$, $\dots$, $c-1$ cannot appear in the
secret, we can replace all of them by $c_0+1$ (here we assume that
$c_0+1 \le c-1$). In other words, we need to consider only $c_0+2$
colors in questions, namely the colors $0$, $1$, $2$, $\dots$,
$c_0+1$, where the number $c_0$ does not depend on $c$, because
$c_0<pr$. This allows us to solve the end-game by the computer
program. We check whether all non-reducible states can be finished in
$q$ questions. If the result is negative, we have the lower bound
$\abb(p,c)>c-r+q$.

The above considerations are taken under the assumption that the
number of colors is sufficiently large. We require that $c$ is the
maximum number of colors used in all checked states, i.e., the maximum
over the values of $c_0+2$ in all states. As we will see later, for a
smaller number of colors some states are impossible. This does not
invalidate the lower bound. Moreover, if we prove a lower bound for a
given state and $c_0+2$ colors, then the lower bound also holds for
the state, when the codebreaker has less than $c_0+2$ colors.
Therefore, we conclude that the lower bound holds for all $c \ge r$.

\begin{figure}[t]
\begin{center}
\begin{displaymath}
A_1=\left(
\begin{array}{ccccc}
  0 & 1 & 2 & 3 & 4 \\
  0 & 1 & 2 & 3 & 4 \\
  0 & 1 & 2 & 3 & 4 \\
\end{array}
\right)\quad
A_2=\left(
\begin{array}{cccccc}
  0 & 1 & 2 & 3 & 4 &   \\
  0 & 1 & 2 & 3 &   & 5 \\
  0 & 1 & 2 &   & 4 & 5 \\
\end{array}
\right)
\end{displaymath}
\begin{displaymath}
A_3=\left(
\begin{array}{ccccccc}
  0 & 1 & 2 & 3 & 4 &   &   \\
  0 & 1 & 2 &   &   & 5 & 6 \\
  0 &   &   & 3 & 4 & 5 & 6 \\
\end{array}
\right)
\end{displaymath}
\caption{The non-reducible states for $p=3$}
\label{fig:abbl:states3}
\end{center}
\end{figure}

\subsubsection{Two Pegs}

For $p=2$ we choose $q=1$ and $r=2$. After applying Rule 1, each row
of the state contains $2$ colors. As $2$ colors with disjoint single
row sets can be merged into $1$ color by Rule 4, we only have pair row
sets. By applying Rule 2, all states are reducible to the single state
$\bigl(
\begin{smallmatrix}
  0 & 1 \\
  0 & 1 \\
\end{smallmatrix}
\bigr)$.
As the state has $2$ secrets, the end-game cannot be won in $1$
question, which implies the lower bound $\abb(2,c)>c-r+q=c-1$ for
$c\ge2$. Therefore, we have shown the inequality ``$\ge$'' in equation
(\ref{eq:abb2}).

\begin{figure}[t]
\begin{center}
\begin{displaymath}
\begin{array}{cc}
B_1=\left(
\begin{array}{ccccc}
  0 & 1 & 2 & 3 & 4 \\
  0 & 1 & 2 & 3 & 4 \\
  0 & 1 & 2 & 3 & 4 \\
  0 & 1 & 2 & 3 & 4 \\
\end{array}
\right)
&
B_2=\left(
\begin{array}{cccccc}
  0 & 1 & 2 & 3 & 4 &   \\
  0 & 1 & 2 & 3 & 4 &   \\
  0 & 1 & 2 & 3 &   & 5 \\
  0 & 1 & 2 &   & 4 & 5 \\
\end{array}
\right)
\\[1cm]
B_3=\left(
\begin{array}{ccccccc}
  0 & 1 & 2 & 3 & 4 &   &   \\
  0 & 1 & 2 & 3 &   & 5 &   \\
  0 & 1 & 2 & 3 &   &   & 6 \\
  0 & 1 &   &   & 4 & 5 & 6 \\
\end{array}
\right)
&
B_4=\left(
\begin{array}{cccccc}
  0 & 1 & 2 & 3 & 4 &   \\
  0 & 1 & 2 & 3 &   & 5 \\
  0 & 1 & 2 &   & 4 & 5 \\
  0 & 1 &   & 3 & 4 & 5 \\
\end{array}
\right)
\\[1cm]
B_5=\left(
\begin{array}{ccccccc}
  0 & 1 & 2 & 3 & 4 &   &   \\
  0 & 1 & 2 & 3 & 4 &   &   \\
  0 & 1 & 2 &   &   & 5 & 6 \\
  0 &   &   & 3 & 4 & 5 & 6 \\
\end{array}
\right)
&
B_6=\left(
\begin{array}{ccccccc}
  0 & 1 & 2 & 3 & 4 &   &   \\
  0 & 1 & 2 & 3 &   & 5 &   \\
  0 & 1 & 2 &   & 4 &   & 6 \\
  0 &   &   & 3 & 4 & 5 & 6 \\
\end{array}
\right)
\\[1cm]
B_7=\left(
\begin{array}{cccccccc}
  0 & 1 & 2 & 3 & 4 &   &   &   \\
  0 & 1 & 2 & 3 &   & 5 &   &   \\
  0 & 1 & 2 &   &   &   & 6 & 7 \\
    &   &   & 3 & 4 & 5 & 6 & 7 \\
\end{array}
\right)
&
B_8=\left(
\begin{array}{ccccccc}
  0 & 1 & 2 & 3 & 4 &   &   \\
  0 & 1 & 2 & 3 &   & 5 &   \\
  0 & 1 &   &   & 4 & 5 & 6 \\
    &   & 2 & 3 & 4 & 5 & 6 \\
\end{array}
\right)
\\[\parskip]
\end{array}
\end{displaymath}
\caption{The non-reducible states for $p=4$}
\label{fig:abbl:states4}
\end{center}
\end{figure}

\subsubsection{Three Pegs}

For $p=3$ we choose $q=r=5$. In the following, we will show that all
states are reducible to the only $3$ ones which are shown in Figure
\ref{fig:abbl:states3}. After applying Rule 1, each row of the state
contains $5$ colors. After that, if the state contains a single row
set, then it must contain another row set which is disjoint with it.
These row sets can be merged by Rule 4. Hence in the following, we
assume that the state does not contain single row sets and the state
table contains exactly $15$ elements. We consider four cases
distinguishing the number of triple row sets in the state.

\begin{itemize}

\item An even number of colors has a triple row set. Then there is an
odd number of remaining elements in the state table. This would mean
that $1$ color has a single row set, and we have a contradiction.

\item $5$ colors have a triple row set. Then after applying Rule
2, we receive table $A_1$ of Figure~\ref{fig:abbl:states3}.

\item $3$ colors have a triple row set. There are $6$ remaining
elements in the state table. There must be $3$ colors, each
having a pair row set. By applying Rules 2 and 3, we receive
table $A_2$ of Figure \ref{fig:abbl:states3}.

\item $1$ color has a triple row set. There are $12$ remaining
elements in the state table. There must be $6$ colors, each
having a pair row set. By applying Rules 2 and 3, we receive
table $A_3$ of Figure~\ref{fig:abbl:states3}.

\end{itemize}

The computer experiment shows that neither of the states $A_1$, $A_2$
and $A_3$ can be solved in $5$ questions, which yields
$\abb(3,c)>c-r+q=c$ for $c \ge 5$. Note that for smaller values of
$c$, some states are impossible in the end-game. Only $A_1$ appears
for $c=5$, only $A_2$ for $c=6$, but for $c \ge 7$ all three states
could appear. The same lower bound for $3 \le c \le 4$ is quite easy
to check directly by the computer program. Hence, we have shown the
inequality ``$\ge$'' in equation~(\ref{eq:abb3}).

\subsubsection{Four Pegs}

For $p=4$ we also choose $q=r=5$. We will show that all states are
reducible to the only $8$ ones which are shown in Figure
\ref{fig:abbl:states4}. First, we apply Rule 1 so that each row of the
state contains exactly $5$ colors. Next, we apply Rule 4 as long as
all disjoint row sets are eliminated. Among others this eliminates all
single row sets. The following observation is easy to see.

\case{Observation}
Consider a state containing only $n$ pairwise non-disjoint pair
row sets. Then there exists an empty row or a row containing $n$
different colors.

Patterns of pairwise non-disjoint pair row sets are shown in Figure
\ref{fig:abbl:patterns2}.

\begin{figure}[t]
\begin{center}
\begin{displaymath}
P_1=\left(
\begin{array}{c}
    \\
    \\
  0 \\
  0 \\
\end{array}
\right)\quad
P_2=\left(
\begin{array}{cc}
    &   \\
  0 &   \\
    & 1 \\
  0 & 1 \\
\end{array}
\right)\quad
P_3=\left(
\begin{array}{ccc}
    &   &   \\
  0 &   & 2 \\
    & 1 & 2 \\
  0 & 1 &   \\
\end{array}
\right)\quad
P_4=\left(
\begin{array}{ccc}
  0 &   &   \\
    & 1 &   \\
    &   & 2 \\
  0 & 1 & 2 \\
\end{array}
\right)
\end{displaymath}
\caption{States with pairwise non-disjoint pair row sets}
\label{fig:abbl:patterns2}
\end{center}
\end{figure}

The observation implies that a state contains at most $4$ colors with
a pair row set, which can be seen as follows. Assume that a state
contains more than $4$ colors with a pair row set. The state table
contains $20$ elements. Thus it contains $1$ color with a quadruple
row set and $2$ colors with a triple row set, or $2$ colors with a
quadruple row set, or $2$ colors with a triple row sets, or $1$ color
with a quadruple row set, or no other colors. Then there exists a row
with at most $3$ colors or at least $6$ colors, which is a
contradiction.

The observation also implies that if a state contains the same number
of colors in each row and a color with a pair row set, it must also
contain a color with a triple row set. We consider six cases
distinguishing the number of quadruple row sets.

\begin{itemize}

\item $5$ colors have a quadruple row set. Then after applying Rule 2,
we receive table $B_1$ of Figure~\ref{fig:abbl:states4}.

\item $4$ colors have a quadruple row set. There are $4$ remaining
elements in the state table. This means that $2$ disjoint pair row
sets exist, which is a contradiction.

\item $3$ colors have a quadruple row set. There are $8$ remaining
elements in the state table. By the second conclusion of the
observation, $2$ colors have a triple row set and $1$ color has a pair
row set. The triple row sets are distinct, as otherwise one row would
contain $6$ colors. By applying Rules 2 and 3, we receive table $B_2$
of Figure \ref{fig:abbl:states4}.

\item $2$ colors have a quadruple row set. There are $12$ remaining
elements in the state table. By the first conclusion of the
observation, we have two sub-cases.

\begin{itemize}

\item $2$ colors have a triple row set and $3$ colors have a pair row
set. If the triple row sets are distinct, then there exists a row with
at most $4$ colors or at least $6$ colors. Therefore, the triple row
sets must be equal. By applying Rules 2 and 3, we receive table $B_3$
of Figure \ref{fig:abbl:states4}.

\item $4$ colors have a triple row set. All triple row sets are
distinct, as otherwise a row would contain at most $4$ colors. By
applying Rules 2 and 3, we receive table $B_4$ of
Figure~\ref{fig:abbl:states4}.

\end{itemize}

\item $1$ color has a quadruple row set. There are $16$ remaining
elements in the state table. By the first conclusion of the
observation, $4$ colors have a triple row set and $2$ colors have a
pair row set. We have to distinguish the relations between the $4$
triple row sets.

\begin{itemize}

\item There are $4$ different triple row sets. Then there exists a
row which contains $4$ or $6$ colors. We have a contradiction.

\item There are $3$ equal triple row sets. Then there exists a row
which contains at most $4$ colors. We have a contradiction.

\item There are exactly $2$ equal triple row sets and $2$ further
equal triple row sets. Then by applying Rules 2 and 3, we receive
table $B_5$ of Figure \ref{fig:abbl:states4}.

\item There are exactly $2$ equal triple row sets and $2$ further
different triple row sets. Then by applying Rules 2 and 3, we
receive table $B_6$ of Figure~\ref{fig:abbl:states4}.

\end{itemize}

\item $0$ colors have a quadruple row set. There are $20$ remaining
elements in the state table. By the first conclusion of the
observation, we have two sub-cases.

\begin{itemize}

\item $4$ colors have a triple row set and $4$ colors have a pair row
set. It is not possible that a row exists which contains no colors of
pair row sets, as otherwise this row would contain not more than $4$
colors. By the observation, a row exists which contains $4$ colors of
pair row sets. Thus this row contains only $1$ color of triple row
sets. This means that $1$ color has a triple row set and further $3$
colors have another equal triple row set. Now, the pair row sets are
uniquely determined. By applying Rules 2 and 3, we receive table $B_7$
of Figure \ref{fig:abbl:states4}.

\item $6$ colors have a triple row set and $1$ color has a pair row
set. We have to distinguish the relations between the $6$ triple row
sets.

\begin{itemize}

\item There are at least $3$ equal triple row sets $R_1$. W.l.o.g.,
let $R_1=\{0,1,2\}$. Then row $3$ contains at most $4$ colors, which
leads to a contradiction.

\item There are $2$ equal triple row sets $R_1$, $2$ further equal
triple row sets $R_2$, and $2$ further equal triple row sets $R_3$.
W.l.o.g., let $R_1=\{0,1,2\}$, $R_2=\{0,1,3\}$, $R_3=\{0,2,3\}$. Then
row $0$ contains at least $6$ colors, which leads to a contradiction.

\item There are $2$ equal triple row sets $R_1$, $2$ further equal
triple row sets $R_2$, and $2$ further different triple row sets
$R_3$, $R_4$. W.l.o.g., let $R_1=\{0,1,2\}$, $R_2=\{0,1,3\}$,
$R_3=\{0,2,3\}$, $R_4=\{1,2,3\}$. Then the pair row set of the
remaining color is uniquely determined as $\{2,3\}$. By applying Rules
2 and 3, we receive table $B_8$ of Figure \ref{fig:abbl:states4}.

\end{itemize}

\end{itemize}

\end{itemize}

The computer experiment shows that neither of the states $B_1$, $B_2$,
$\dots$, $B_8$ can be solved in $5$ questions, which yields
$\abb(4,c)>c-r+q=c$ for $c \ge 5$. Again for smaller values of $c$,
some states are impossible, e.g., for $c=7$ state $B_7$ cannot appear.
The same lower bound for $c=4$ is quite easy to check directly by the
computer program. Hence, we have shown the
inequality~(\ref{eq:abb4lower}).

\subsection{Upper Bounds}

We prove upper bounds of the $\ABB(p,c)$ game by showing a strategy
for the codebreaker. Questions of the following form will play a major
role in the strategy:
\begin{displaymath}
\mq{k \bmod c, \; k+1 \bmod c, \; \dots, \; k+p-1 \bmod c}
\end{displaymath}
for a given $k \in \mathbb{N} $ (not necessarily in
$\{0,1,\dots,c-1\}$). We will denote such a question by $\mq{k}$ for
short. The strategy consists of two phases: the reduction and the
end-game.

\case{The reduction}
The codebreaker starts with the question $\mq{0}$ and asks totally at
most $x$ questions. He or she follows three rules.

\subcase{Rule 1}
If the codemaker answers with $p$ black pegs, the game is finished.

\subcase{Rule 2}
As long as the codemaker answers with zero black pegs, the
codebreaker continues with consecutive questions in decreasing order:
$\mq{c-1}$, $\mq{c-2}$, $\mq{c-3}$, etc.

\subcase{Rule 3}
If question $\mq{k}$ is the first one answered with $b$ black pegs,
where $1 \le b \le p-1$, the codebreaker begins to ask questions in
increasing order, i.e., instead of asking $\mq{k-1}$, he or she
asks questions $\mq{1}$, $\mq{2}$, etc., as next.

\case{The end-game}
After the $x$ questions of the reduction phase, if the game has not
yet been finished, the codebreaker plays using all possible questions.

This two phase strategy is based on three ideas. First, for a given
fixed number of pegs all end-games with an arbitrary large number of
colors can be reduced to an end-game with a finite and small number of
colors. Second, the end-game can be effectively solved by a variant of
the computer program. Third, the Rule 3 is substantial. Without it the
tight upper bound cannot be obtained.

\begin{figure}[t]
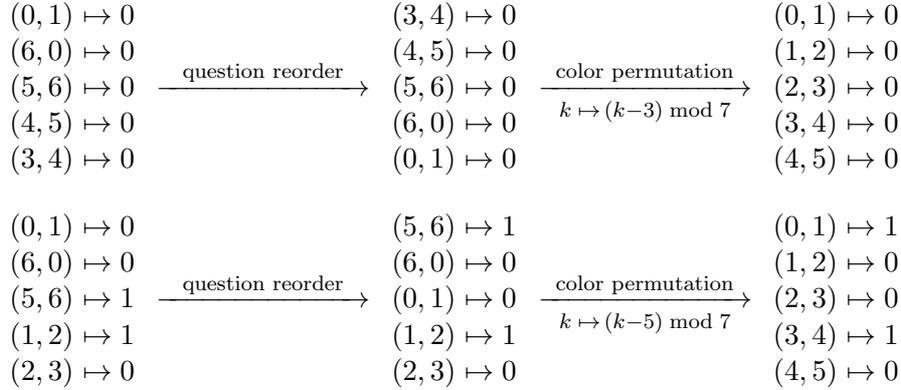

\begin{center}
\begin{displaymath}
\begin{array}{c}
  (0,1) \mapsto 0 \\
  (6,0) \mapsto 0 \\
  (5,6) \mapsto 0 \\
  (4,5) \mapsto 0 \\
  (3,4) \mapsto 0 \\
\end{array}
\mathrel{\mathop{\kern0pt{\hbox to80pt{\rightarrowfill}}}
\limits^{\textnormal{question reorder}}}
\begin{array}{c}
  (3,4) \mapsto 0 \\
  (4,5) \mapsto 0 \\
  (5,6) \mapsto 0 \\
  (6,0) \mapsto 0 \\
  (0,1) \mapsto 0 \\
\end{array}
\mathrel{\mathop{\kern0pt{\hbox to80pt{\rightarrowfill}}}
\limits^{\textnormal{color permutation}}
\limits_{k\;\mapsto\,(k-3)\bmod7}}
\begin{array}{c}
  (0,1) \mapsto 0 \\
  (1,2) \mapsto 0 \\
  (2,3) \mapsto 0 \\
  (3,4) \mapsto 0 \\
  (4,5) \mapsto 0 \\
\end{array}
\end{displaymath}
\begin{displaymath}
\begin{array}{c}
  (0,1) \mapsto 0 \\
  (6,0) \mapsto 0 \\
  (5,6) \mapsto 1 \\
  (1,2) \mapsto 1 \\
  (2,3) \mapsto 0 \\
\end{array}
\mathrel{\mathop{\kern0pt{\hbox to80pt{\rightarrowfill}}}
\limits^{\textnormal{question reorder}}}
\begin{array}{c}
  (5,6) \mapsto 1 \\
  (6,0) \mapsto 0 \\
  (0,1) \mapsto 0 \\
  (1,2) \mapsto 1 \\
  (2,3) \mapsto 0 \\
\end{array}
\mathrel{\mathop{\kern0pt{\hbox to80pt{\rightarrowfill}}}
\limits^{\textnormal{color permutation}}
\limits_{k\;\mapsto\,(k-5)\bmod7}}
\begin{array}{c}
  (0,1) \mapsto 1 \\
  (1,2) \mapsto 0 \\
  (2,3) \mapsto 0 \\
  (3,4) \mapsto 1 \\
  (4,5) \mapsto 0 \\
\end{array}
\end{displaymath}
\caption{Game state examples for $p=2$, $c=7$, $x=5$}
\label{fig:abbu:reduct_upp}
\end{center}
\end{figure}

\begin{figure}[t]
\begin{center}
\begin{displaymath}
\begin{array}{c}
  (0,1) \mapsto 0 \\
  (1,2) \mapsto 0 \\
  (2,3) \mapsto 0 \\
  (3,4) \mapsto 0 \\
  (4,5) \mapsto 0 \\
\end{array}
\mathrel{\mathop{\kern0pt{\hbox to80pt{\rightarrowfill}}}
\limits^{\textnormal{color mapping}}
\limits_{\substack{
  0 \;\mapsto\, 0 \\
  5 \;\mapsto\, 3 \\
  6 \;\mapsto\, 4
}}}
\begin{array}{c}
  (0,1) \mapsto 0 \\
  (1,2) \mapsto 0 \\
  (2,3) \mapsto 0 \\
\end{array}
\end{displaymath}
\begin{displaymath}
\begin{array}{c}
  (0,1) \mapsto 1 \\
  (1,2) \mapsto 0 \\
  (2,3) \mapsto 0 \\
  (3,4) \mapsto 0 \\
  (4,5) \mapsto 0 \\
\end{array}
\mathrel{\mathop{\kern0pt{\hbox to80pt{\rightarrowfill}}}
\limits^{\textnormal{color mapping}}
\limits_{\substack{
  0 \;\mapsto\, 0 \\
  1 \;\mapsto\, 1 \\
  5 \;\mapsto\, 3 \\
  6 \;\mapsto\, 4
}}}
\begin{array}{c}
  (0,1) \mapsto 1 \\
  (1,2) \mapsto 0 \\
  (2,3) \mapsto 0 \\
\end{array}
\end{displaymath}
\begin{displaymath}
\begin{array}{c}
  (0,1) \mapsto 1 \\
  (1,2) \mapsto 0 \\
  (2,3) \mapsto 0 \\
  (3,4) \mapsto 1 \\
  (4,5) \mapsto 0 \\
\end{array}
\mathrel{\mathop{\kern0pt{\hbox to80pt{\rightarrowfill}}}
\limits^{\textnormal{color mapping}}
\limits_{\substack{
  0 \;\mapsto\, 0 \\
  1 \;\mapsto\, 1 \\
  3 \;\mapsto\, 2 \\
  4 \;\mapsto\, 3
}}}
\begin{array}{c}
  (0,1) \mapsto 1 \\
  (1,2) \mapsto 0 \\
  (2,3) \mapsto 1 \\
\end{array}
\end{displaymath}
\caption{Color mapping examples for $p=2$, $c_1=7$, $x_1=5$, $c_0=5$,
$x_0=3$}
\label{fig:abbu:mapping}
\end{center}
\end{figure}

The state of the game after the reduction phase is uniquely determined
by the set of pairs ``question---answer'', where the order of the
questions is not important. Moreover, if we permute the colors, the
worst case number of questions remains unchanged. In our case, it
suffices to rotate the colors. Hence, we can restrict our
considerations to the sequence of questions $\mq{0}, \mq{1}, \dots,
\mq{x-1}$, where either all answers are zero black pegs or the
first asked question $\mq{0}$ is answered with at least one black peg.
Two examples are shown in Figure \ref{fig:abbu:reduct_upp}. In the top
example, all questions are answered with zero black pegs. In the
bottom example, some questions are answered with a non-zero number of
black pegs.

Observe that in the reduction phase, if the number of colors is large,
the most questions are answered with zero black pegs. In fact, only at
most $p$ questions can get another answer. As a color used at a given
position in a question answered with zero black pegs cannot appear at
this position in the secret, after the reduction phase the most colors
are excluded from being in the secret.

\begin{figure}[t]
\begin{center}
\begin{displaymath}
\begin{array}{ccc}
(0,0,0,0,0,0,0) & (1,0,0,0,0,0,0) & (2,0,0,0,0,0,0) \\
(1,1,0,0,0,0,0) & (1,0,1,0,0,0,0) & (1,0,0,1,0,0,0) \\
(1,0,0,0,1,0,0) & (1,0,0,0,0,1,0) & (1,0,0,0,0,0,1) \\
(1,2,0,0,0,0,0) & (1,0,2,0,0,0,0) & (1,0,0,2,0,0,0) \\
(1,0,0,0,2,0,0) & (1,0,0,0,0,2,0) & (1,0,0,0,0,0,2) \\
(2,1,0,0,0,0,0) & (2,0,1,0,0,0,0) & (2,0,0,1,0,0,0) \\
(2,0,0,0,1,0,0) & (2,0,0,0,0,1,0) & (2,0,0,0,0,0,1) \\
(1,1,1,0,0,0,0) & (1,1,0,1,0,0,0) & (1,1,0,0,1,0,0) \\
(1,1,0,0,0,1,0) & (1,1,0,0,0,0,1) & (1,0,1,1,0,0,0) \\
(1,0,1,0,1,0,0) & (1,0,1,0,0,1,0) & (1,0,1,0,0,0,1) \\
(1,0,0,1,1,0,0) & (1,0,0,1,0,1,0) & (1,0,0,1,0,0,1) \\
(1,0,0,0,1,1,0) & (1,0,0,0,1,0,1) & (1,0,0,0,0,1,1) \\
\end{array}
\end{displaymath}
\caption{Sequences of answers for the end-game with $p=3$}
\label{fig:abbu:36}
\end{center}
\end{figure}

To be more formally, consider two games $\ABB(p,c_1)$ and
$\ABB(p,c_0)$, where $c_1 \ge c_0$. Let the number of questions in the
reduction phase be $x_1=c_1-y$ and $x_0=c_0-y$, respectively, where $y
\in \mathbb{N}$ with $y \le c_0$. We want to use the strategy of the
$\ABB(p,c_0)$ end-game in the $\ABB(p,c_1)$ end-game, which for
$c_1=c_0$ are obviously the same strategies. The idea relies on color
mapping, which must take into account all colors not excluded in the
reduction phase, and only these colors. In particular, we should
consider $p$ questions with a pairwise disjoint set of colors and each
answered with one black peg. Hence, we must additionally assume
that $c_0 \ge p^2$ and $x_0 \ge p^2-p+1$. Examples are shown in Figure
\ref{fig:abbu:mapping}. The left column contains the questions and
answers after the reduction phase of $\ABB(p,c_1)$. The right column
contains the questions and answers after the reduction phase of
$\ABB(p,c_0)$. The examples cover three important situations. In the
top example, all answers are zero black pegs. The colors $5$ and
$6$ are allowed at position $0$, and the colors~$0$ and $6$ at
position $1$. In the middle example, at least one answer received a
non-zero number of black pegs, but the sum of received black pegs is
less than the number of pegs. The colors $0$, $5$ and $6$ are allowed
at position $0$, and the colors $0$, $1$ and $6$ at position $1$. In
the bottom example, the sum of received black pegs is equal to the
number of pegs. The colors $0$ and $3$ are allowed at position $0$,
and the colors $1$ and~$4$ at position~$1$. Finally, if there exists a
$q$ such that we find a winning strategy for every $\ABB(p,c_0)$
end-game in at most $q-x_0$ questions, then the end-strategy is
applicable to the $\ABB(p,c_1)$ end-game as well. Therefore,
$\abb(p,c_1) \le q-x_0+x_1 = q+c_1-c_0$.

Now, to prove the upper bound for a given $p$, we choose appropriate
values of $c$, $q$ and $x$, where
\begin{equation}
\label{eq:abbu:assumption}
c \ge p^2 \quad\textnormal{and}\quad x \ge p^2-p+1.
\end{equation}
We check by the computer program whether the $\ABB(p,c)$ end-game can
be finished in $q-x$ questions. Let $(b_1,b_2,\dots,b_{x})$ be a
sequence of answers in the reduction phase. As argued above, we have
to consider only sequences of answers, where either
$b_1=b_2=\dots=b_{x}=0$ or $b_1\ne0$.

\subsubsection{Two Pegs}

For $p=2$ we choose $c=q=5$ and $x=3$. There are four sequences of
answers: $(0,0,0)$, $(1,0,0)$, $(1,1,0)$, $(1,0,1)$.
The computer experiment shows that all four end-games can be finished
in $q-x=2$ questions, which yields the desired $c$-question upper
bound (i.e., $\abb(2,c) \le c$) for $c \ge 5$. The computer program
finds a $c$-question strategy for $c=2,3,4$, which can also be easily
checked by hand. Hence, we have shown the inequality ``$\le$'' in
equation (\ref{eq:abb2}).

\subsubsection{Three Pegs}

For $p=3$ we choose $c=9$, $q=10$ and $x=7$. The $36$ possible
sequences of answers are shown in Figure \ref{fig:abbu:36}. By further
symmetries they can be reduced to only $17$ ones, shown in Figure
\ref{fig:abbu:17}. The computer experiment shows that all end-games
are finished in $q-x=3$ questions, which yields the $(c+1)$-question
upper bound for $c \ge 9$. The same upper bound for $3 \le c \le 8$
can be quite easy checked by the computer program. Hence, we have
shown the inequality ``$\le$'' in equation (\ref{eq:abb3}).

\begin{figure}[t]
\begin{center}
\begin{displaymath}
\begin{array}{ccc}
(0,0,0,0,0,0,0) & (1,0,0,0,0,0,0) & (2,0,0,0,0,0,0) \\
(1,1,0,0,0,0,0) & (1,0,1,0,0,0,0) & (1,0,0,1,0,0,0) \\
(1,0,0,0,0,1,0) & (1,0,0,0,0,0,1) & (1,1,1,0,0,0,0) \\
(1,1,0,1,0,0,0) & (1,1,0,0,1,0,0) & (1,0,1,0,1,0,0) \\
(1,0,1,0,0,1,0) & (1,0,0,1,0,0,1) & (2,1,0,0,0,0,0) \\
(2,0,1,0,0,0,0) & (2,0,0,1,0,0,0) \\
\end{array}
\end{displaymath}
\caption{Non-isomorphic sequences of answers for the end-game
with $p=3$}
\label{fig:abbu:17}
\end{center}
\end{figure}

\subsubsection{Four Pegs}

For $p=4$ we choose $c=16$, $q=18$ and $x=13$. There are $560$
sequences of answers, which are reducible to only $117$. However,
there are still too many cases to be presented here. The computer
experiment shows that all end-games are finished in $q-x=5$ questions,
which yields the $(c+2)$-question upper bound for $c \ge 16$.

An optimal $(c+1)$-question strategy for $4 \le c \le 7$ can be easily
found in a few seconds by the computer program. The cases $8 \le c \le
15$ need some more effort. To reduce computation time we search only
for two phase strategies. Note that we omit the assumptions
(\ref{eq:abbu:assumption}), because we want only a strategy for a
fixed number of colors. For $c=8,9,10$ we apply $q=c+1$ and $x=c-4$.
We receive $35$, $56$, $84$ cases, respectively. For $11 \le c \le 15$
we apply $q=c+2$ and $x=c-3$. We receive $165$, $220$, $286$, $364$,
$455$ cases, respectively. Some of the cases are isomorphic. However,
the time spending on eliminating isomorphisms would be longer than the
time needed to solve all cases. Therefore, we omit this step. And
again all end-games finish in $q-x=5$ questions, which finally
confirms that we have shown inequality~(\ref{eq:abb4upper}).

\section{Conclusions and Future Work}\label{conclusions}

In this paper we have proved exact values for $\ab(2,c)$, $\ab(3,c)$,
$\abb(2,c)$, $\abb(3,c)$, and tight bounds for $\ab(4,c)$ and
$\abb(4,c)$. These proofs for $p=2$, $3$, $4$ are all based on the
idea of reducing the game with an arbitrary number of colors to a game
with a small number of colors and solving it by computer. This idea is
general and may be applicable for any constant number of pegs.
However, there are two problems, namely generating the growing number
of end-games and solving all these end-games. The latter problem seems
to be computationally harder and requires new ideas, as the end-games
need to be played with approximately $p^2$ colors. This is still too
much for $p \ge 4$, as the number of possible secrets (and thus the
computational complexity) increases asymptotically like $c^p$.
Another interesting case is the game with equal number of pegs and
colors, where the AB game and the ABB game equal. For this case we
proved only a lower bound. We need new ideas here, as the strategies
leading to the values for a fixed number of pegs do not seem to be
well applicable for it.

Looking at the presented results, one can conjecture that for the
$\AB(p,c)$ game the number of questions in the worst case grows like
the fraction $c/p$, but for the $\ABB(p,c)$ game with at least $3$
pegs it seems to be independent of the number of pegs and to be equal
to $c+1$. Note that if the formula $\ABB(p,c) = c+1$ for $p \ge 3$
could be proved, we would have a complete formula for the Generalized
Black-peg AB game. This would be rather interesting, as this game
would not become more difficult for increasing $p$.

Further work should concentrate on closing the gap between lower and
upper bounds for $4$ pegs, on the case of $5$ pegs and on the case of
equal number of pegs and colors.

\vspace*{-0.5em}
% Gerold I do not want to have an extra page only for one citation.

\end{document}